\documentclass[12pt,preprint]{aastex}

\slugcomment{To appear in Astrophysical Journal, 2007, Vol. 657, No. 1}

\shorttitle{CO line width and black hole - bulge relationship}
\shortauthors{Wu}

\begin{document}
\title{CO line width and the black hole -- bulge relationship 
at high redshift}


\author{Xue-Bing Wu} 
\affil {Department of Astronomy, Peking University, Beijing 
100871, China}

\email{wuxb@bac.pku.edu.cn}

\begin{abstract}

Recently, it has been suggested that the CO line width (FWHM(CO)) is a surrogate 
for the bulge velocity dispersion ($\sigma$) of the host galaxies of high-redshift 
quasars, and the black hole --
bulge ($M_{BH}-\sigma$) relation obtained with this 
assumption departs significantly from the $M_{BH}-\sigma$ relation in the 
local universe. In this study, we first present an investigation of the 
correlation
between the CO line width and the bulge velocity dispersion using a sample of
33 nearby Seyfert galaxies. We find that the formula adopted in previous 
studies, $\sigma=\rm{FWHM(CO)}/2.35$, is generally not a good approximation. 
Using it, one may
underestimate the value of bulge velocity dispersion significantly when the 
CO line is narrower
than 400~$km~s^{-1}$. By involving the galactic inclination angle $i$ as an
additional parameter, we obtain a tight correlation between the 
inclination-corrected CO line width
 and the bulge velocity dispersion, namely,
$\rm {FWHM(CO)}/\sin i=-67.16\pm80.18+(3.62\pm0.68)\sigma$. 
Using this new relation, we can better
estimate the bulge velocity dispersion from the CO line width if the galactic
inclination is known. We apply this new relation to nine high-redshift quasars 
with CO line detections and find that
they are consistent with the local $M_{BH}-\sigma$ relation if their 
inclination angles are around $15^o$. The possible smaller inclinations of the
high-redshift quasars are preferred because of their relatively greater likelihood 
of detection, 
and are also 
consistent with their relatively smaller CO line
widths compared to  submillimeter galaxies (SMGs) at high redshift
having a similar total amount of molecular gas. Future observations
are needed to confirm these results.

\end{abstract}


\section{Introduction}

The masses of supermassive black holes in nearby normal and active
galaxies have been reliably derived in the last two decades by using different 
dynamical techniques (Kormendy \& Gebhardt
2001; Ho 1999). The relationships of the black hole mass to the bulge 
velocity dispersion and to the bulge mass have been established for both
quiescent and active galaxies (Gebhardt et al. 2000; Ferrerase et al. 2001;
Tremaine et al. 2002; Onken et al. 2004). Various
theoretical models have been suggested to explain the tight black hole -- bulge
relations (Silk \& Rees 1998; King 2003, 2005; Wyithe \& Loeb 2003; 
Murray, Quataert
 \& Thompson 2005; Begelman \& Nath 2005). Most models and simulations
predict that the  
$M_{BH}-\sigma$ relation may be the same at different redshifts (King 2003, 2005
Wyithe \& Loeb 2003; Begelman \& Nath 2005; Robertson et al. 2006), but some 
do not (Kazantzidis et al. 2005). On the observational side, 
testing the $M_{BH}-\sigma$ relation
at high redshift is a great challenge. Although we do have some methods to 
estimate the 
black hole masses of high-redshift quasars, especially by measuring their
broad emission line properties (McLure \& Jarvis 2002; Vestergaard 2002; Kong et al.
2005), 
it is difficult to measure the bulge
properties in the host galaxies of these quasars directly. 

Even for quasars at low redshift, it is still not easy to resolve their host 
galaxies observationally
because their central nuclei are too bright. Measuring the velocity dispersion
of their host galaxies with spectroscopic methods is also difficult.  Only 
very recently were the near-infrared measurements of the bulge 
velocity dispersions obtained for 11 bright
Palomar-Green (PG) quasars (Dasyra et al. 2006).
However, for local quasars, one can use the 
empirical relation between the [OIII] line width and bulge velocity dispersion,
$\sigma=$FWHM([OIII])/2.35, derived from local Seyfert galaxies 
(Nelson \& Whittle 1996), to estimate their bulge velocity dispersions and then
study the  $M_{BH}-\sigma$ relation for local active galactic nuclei (AGNs; Shields et al. 2003). 
But such an approach
can not be applied to quasars at high redshift, because their [OIII] lines move
out of the optical windows and are not observable. 

Recently, Shields et al. 
(2006) suggested using the CO line width as a surrogate for the bulge velocity 
dispersion. CO lines have been clearly detected in about 10 quasars with 
redshift from 1 to 6.4 (Solomon \& Vanden Bout 2005). 
Their CO line widths are around several hundred 
$km~s^{-1}$, comparable to the typical width of narrow emission 
lines of AGNs. By adopting the formula $\sigma=$FWHM(CO)/2.35, Shields et al.
 (2006)
investigated the $M_{BH}-\sigma$ relation in high-redshift quasars and found that
it departs significantl from the local $M_{BH}-\sigma$ relation. 
The same black hole mass may correspond to a smaller $\sigma$ value at higher
redshift.
However,
whether or not the CO line width can be taken as a surrogate for the bulge
velocity dispersion obviously deserves further study. As noted by  Shields et al.
 (2006), the relatively narrower CO line width of high-redshift quasars may
be due to their smaller inclinations. Face-on quasars could be preferentially
found at higher redshift than quasars with larger inclinations. Clearly, more
works on the CO line width and the black hole -- bulge relation of high-redshift 
quasars are expected.

Fortunately, the CO line width and bulge velocity dispersion have been obtained for
some nearby Seyfert galaxies. An early survey by Heckman et al. (1989) detected
 CO(1--0) emission in 18 
out of 43 Seyfert galaxies using the NRAO 12 m telescope. With the same 
instrument, Maiolino et al. 
(1997) successfully detected CO(1--0) emission in 73 out of 94 Seyfert 
galaxies in  
the Revised Shapely Ames (RSA) and Center for Astrophysics (CfA) Seyfert 
samples (Maiolino \& Rieke 1995; Huchra \& Burg 1992). 
For the bulge velocity dispersion, Nelson \& Whittle (1995) have compiled
a catalog of the derived values for 78 Seyfert galaxies. Therefore, we
can use the Seyfert galaxies with the published data for both the CO line width
and the bulge velocity dispersion to study the relationship between them. 
In fact,
Heckman et al. (1989) have investigated such a relation in seven Seyfert galaxies
(see their Table 8 and Figure 13) and found that the inclination-corrected
CO line width may correlate better with the bulge velocity
dispersion. Obviously, this
tentative result needs to be confirmed by a larger sample of Seyfert galaxies.

In this paper, we first present an investigation of the relation
between the inclination-corrected CO line width and the bulge velocity 
dispersion for a relatively larger sample of Seyfert galaxies. Based
on the new relation derived by us, we estimate the bulge velocity 
dispersions
for high redshift quasars and study whether or not the $M_{BH}-\sigma$
relation at high redshift is different from the local one. Finally we
give our summary and discussions.

\section{CO line width and velocity dispersion of Seyfert galaxies}

CO lines have been successfully detected in a number of Seyfert galaxies
(Heckman et al. 1989; Maiolino et al. 1997; Vila-Vilaro, Taniguchi \& 
Nakai 1998; Curran,
Aalto \& Booth 2000). To avoid the selection bias, we use the Seyfert sample
in Maiolino et al. (1997), for which the sources were selected from the RSA and CfA 
surveys. Of 73 CO-detected Seyfert galaxies in this sample, we find 33 have
measured bulge velocity dispersion data, either from the catalog of
Nelson \& Whittle (1995) or from the Hyperleda database\footnote{see
http://leda.univ-lyon1.fr} (Paturel et al. 2003). We use this sample of 33 Seyfert galaxies
(hereafter called the main sample) to investigate the relation between the CO 
line width and the bulge velocity
dispersion for AGNs. We also find eight additional AGNs, including
four Seyfert galaxies in Heckman et al. (1989) and four radio galaxies in Evans et al. 
(2005), both with measured data for the CO line width and the bulge
velocity dispersion. 
These eight sources are not  included in the statistical studies, but will be
used to check our results. The data of 33 Seyfert galaxies in our main sample and eight 
additional sources are listed in Table 1. The galactic disk inclination 
angles of these AGNs can be obtained from the Hyperleda database. However,
for NGC 6814 we take an inclination of 20$^o$ from Whittle (1992), rather than the value
82$^o$ given by Hyperleda because  it seems more likely to 
be a face-on galaxy, according to its infrared image (Mulchaey, Regan \& Kundu 1997). 
For other sources, the inclination data listed in 
Whittle (1992) are generally consistent with those given by Hyperleda. 

First, we directly investigate the relation between the CO line width 
and the bulge stellar velocity dispersion for Seyfert galaxies. 
In Figure 1 we plot such a relation for 33 Seyferts in our main sample 
and 8 additional sources. Clearly, there is a good correlation between them. 
For 33 Seyferts in our main sample, the Spearman correlation
coefficient is 0.63, and the null probability is 8.95$\times 10^{-5}$. 
Using the Ordinary 
Least Square (OLS) bisector  method (Isobe et al. 1990), we obtain a 
fitted relation,
\begin{equation}
\rm {FWHM(CO)}=-90.66\pm68.95+(2.90\pm0.58)\sigma .
\end{equation}
Comparing the relation $\sigma$=FWHM(CO)/2.35 with the fitted 
line and the observational data for
Seyferts, we find that, although this previously suggested relation 
is not a bad approximation for the 
correlation, by using it
one can underestimate the $\sigma$ 
value for objects with FWHM(CO) less than 400~$km~s^{-1}$ and overestimate
the $\sigma$ value for objects with broader
CO lines.

In a previous study, Heckman et al. (1989) mentioned that the CO line 
width correlates well with
the galactic inclination, and when the CO line width is corrected
for inclination, its correlation with the bulge velocity dispersion can 
be improved. With an enlarged
sample of 33 Seyferts, we can re-investigate the effect of inclination. In 
Fig. 2 we  plot the
FWHM(CO)/$\sigma$ versus the sine of the inclination. Apparently, 
there is a correlation between them. The Spearman correlation coefficient
is 0.60 and the null probability is 2.28$\times 10^{-4}$. The OLS bisector
fit gives
\begin{equation}
\rm {FWHM(CO)}/\sigma=-0.58\pm0.30+(3.95\pm0.44)\sin i .
\end{equation}
From Fig. 2 we can clearly see that the relation $\sigma$=FWHM(CO)/2.35 is 
generally not
a good approximation. Using it one can significantly underestimate $\sigma$
from the CO line width for the AGNs with smaller inclination, and overestimate
 $\sigma$ for AGNs with higher inclination. Only for galaxies with an 
inclination of about 45$^o$ ($\sin i$ around 0.7), does the relation 
$\sigma$=FWHM(CO)/2.35 apply.  From Fig. 2 we can also see that the
galactic inclinations of both Seyfert 1 and Seyfert 2 galaxies seem to span a
similar range. However, we do see that there are relatively more Seyfert 2 galaxies 
with inclinations larger than 70$^o$ than Seyfert 1 galaxies (see also Table 1). In
our main sample three Seyfert  and one Seyfert 1 galaxy have inclinations larger 
than 70$^o$. This Seyfert 1, NGC 2992, is actually an edge-on Seyfert 1.9 
galaxy. We also note that two Seyfert 2 galaxies with the smallest inclinations in our 
sample, NGC 5929 and NGC 1068, are both intrinsically Seyfert 1 galaxies with 
polarized broad emission lines (Gu \& Huang 2002). Considering  the 
possible selection effect in our incomplete CO-detected Seyfert sample and 
also the complicated relation between the inclinations of a central accretion 
disk and  host galaxy, the apparent distributions of galactic inclinations 
of Seyfert galaxies may not be too hard to understand.   

Finally, we investigate the relation between the inclination-corrected CO
line width (FWHM(CO)/$\sin i$) and the bulge velocity dispersion. Using
seven Seyfert galaxies, Heckman et al. (1989) previously obtained 
$\rm {FWHM(CO)}/\sin i=-20+3.57\sigma$. In Fig. 3 we plot such a relation for
33 Seyferts in our main sample and 8 additional sources. 
For 33 Seyferts in our main sample, the Spearman correlation coefficient is
0.71, and the null probability is 3.18$\times 10^{-6}$. Comparing with
the correlation between FWHM(CO) and $\sigma$, the correlation is improved
when the CO line width is corrected for inclination.  
For 33 Seyferts in our main
sample, we obtain an OLS bisector fit as
\begin{equation}
\rm {FWHM(CO)}/\sin i=-67.16\pm80.18+(3.62\pm0.68)\sigma.
\end{equation}
We note that this relation is in good agreement with the previous result
obtained from a smaller Seyfert sample (Heckman et al. 1989). From Fig. 3 we
see that the inclination-corrected CO line width for both the Seyferts in our 
main sample and eight additional AGNs correlates well with the
bulge velocity dispersion, with only a few outliers. The most extreme 
outlier in Fig. 3 is the radio galaxy 3C 84 (Perseus A), which has a narrower 
CO line width 
(200~$km~s^{-1}$)  relative to its bulge velocity dispersion value 
(272~$km~s^{-1}$). After a close look at the CO(1--0) spectrum of 3C 84 in Fig. 1 of
Evans et al. (2005), we note that its narrower CO line width was probably
obtained by measuring the central peak profile only. There are also clear
subpeaks 
in both the red and blue sides around the central peak. Therefore, we guess 
that the CO line width of 3C 84 may be significantly underestimated. This 
obviously needs to be confirmed by more accurate observations in the future.

The reason for considering the inclination correction of the CO line 
width when investigating its relation to the bulge velocity dispersion is
straightforward. As indicated by Hackman et al. (1989), unlike the [OIII]
lines in the narrow-line region, CO gas is probably arrayed
in a rotating disk that is coplanar with the galaxy disk. When we view the CO
gas with an inclination, what we actually measure is its projected velocity.
Therefore, the inclination-corrected CO line width may better represent the 
intrinsic CO velocity. Therefore, it is not surprising that such an intrinsic CO velocity
has a better correlation with the stellar velocity dispersion in the central
bulge than the observed CO line width.

\section{Application to high-redshift quasars}

CO lines have been detected for some high redshift quasars (Solomon \& 
Vanden Bout 2005) and therefore offer us an opportunity to study the
black hole -- bulge relation at high redshift. Using the UV/optical broad
emission line and continuum properties, Shields et al. (2006) have estimated
the black hole mass of these high-redshift quasars with 
the methods suggested in some previous studies (Kaspi et al. 2000, 2005; 
Vestergaard 2002; McLure \& Dunlop 2004). Taking
the CO line width as a surrogate for the bulge velocity dispersion 
($\sigma$=FWHM(CO)/2.35), Shields et al. (2006) found that these high-redshift
quasars evidently depart from the usual $M_{\rm{BH}}-\sigma$ relation in the local
universe and thus suggested that the giant black holes at high redshift reside in 
the undersized bulge. As we know from our investigations above,
 the observed CO line width needs to be corrected for inclination. In this section,
we use the tightest relation (Eq. 3) we found for Seyfert galaxies in the
section above to 
re-investigate the black hole -- bulge relation at high redshift. 

 We note that the host galaxy morphologies of local Seyfert galaxies and
high-redshift quasars may not be the same. Indeed, the host galaxies of 
Seyfert galaxies are typically spiral, while those of high-redshift quasars 
are rather
complex and usually disturbed. However, there also appears to be spiral 
structure in the central part of host galaxies of high-redshift quasars 
(Hutchings 2004). On the other hand, observations show that the host galaxy 
morphologies of Seyfert galaxies are mostly asymmetric, probably
related to the star-forming activities (Maiolino et al. 1997). Similarly,
the complex 
host galaxy morphologies of high-redshift quasars are probably also due to
the triggers of AGNs and star-forming activities. The same as
Seyfert galaxies, high-redshift quasars are mostly radio-quiet, and their 
masses of total molecular gas are also mainly in the range 10$^9M_\odot$ - 
10$^{11}M_\odot$ (Solomon \& Vanden Bout 2005). Therefore, based on these 
points, we tentatively assume that
the CO dynamics in high-redshift quasars is not dramatically different from 
the local Seyfert galaxies and apply our derived relation between the 
inclination-corrected CO line width and bulge velocity dispersion 
to high-redshift quasars. Obviously, this assumption needs to be 
confirmed by further studies.

Shields et al. (2006) have listed the observed CO line width and the estimated
black hole mass  for nine high-redshift quasars with CO detection in their 
Table 2. As there is no information about the inclination of the host galaxy 
of these high-redshift quasars, it is difficult to derive the 
inclination-corrected CO line
width for them. However,  one usually believes that a quasar at high redshift
 should be more easily 
 detected if its galaxy disk is face-on. Recently, Greve et al. (2005) and
Carilli \& Wang (2006) have found the clear difference in the CO line widths
of quasar host galaxies and SMGs at high redshift.
The CO line widths of high-redshift
quasars are typically a factor of 2.3 narrower than those of other SMGs
with the same amount of molecular gas.  Carilli \& Wang (2006) argued that 
the difference in the
CO line width distribution of high-redshift quasars and SMGs can be explained
if the average inclination of quasars is about 3 times smaller than that of
SMGs.

Assuming that the average galaxy inclination of nine high-redshift quasars with 
CO detections is 15$^o$, we can use Eq. (3) to derive the bulge velocity
dispersion from the CO line width and then investigate the 
$M_{\rm{BH}}-\sigma$ relation for high-redshift quasars. Our result is shown
in Fig. 4, from which we can clearly see that if we use the 
inclination-corrected CO line width to estimate the bulge velocity dispersion,
the $M_{\rm{BH}}-\sigma$ relation for high-redshift quasars is generally
consistent with the local one. In fact, we find that the data point for each 
quasar can match the local $M_{\rm{BH}}-\sigma$ relation very well if we
assume the inclination in the range 10$^o$-30$^o$. 
Obviously, our result is different from that recently obtained by 
Shields et al. 
(2006), who used  $\sigma$=FWHM(CO)/2.35 to estimate the bulge velocity 
dispersions for high-redshift quasars. Therefore, we think it is still 
premature to claim that the black hole -- bulge relationship at high redshift
 is different from the local one. The giant black hole masses of these
high-redshift quasars relative to their
narrower CO line width can be explained by assuming smaller inclinations. If
the inclinations of these quasars are really smaller, the bulge velocity
dispersion values derived from the inclination-corrected CO line widths will
be in the range  250~$km~s^{-1}$ - 500~$km~s^{-1}$, making them consistent with
the local $M_{\rm{BH}}-\sigma$ relation. 

Shields et al. (2006) mention that most of the low-redshift PG quasars seem to follow
the usual $M_{\rm{BH}}-\sigma$ relation, even when using $\sigma$=FWHM(CO)/2.35.  
This can be understood if these quasars have inclinations in a range between 30$^o$ and
45$^o$. From Fig. 2 and Eq. (3), we can see that only if the inclinations are in
such a range is the formula $\sigma$=FWHM(CO)/2.35 not a bad approximation, so the 
 $\sigma$ value derived with this formula may not deviate too much  from the 
value derived by using Eq. (3). However, we noted that even in the low-redshift PG quasar 
sample there are also 
several narrow CO quasars, with CO line width in the range 50~$km~s^{-1}$ - 90 $km~s^{-1}$ 
(Evans et al.
 2001, 2006; Shields et al. 2006). These objects also seem to be  outliers of the 
$M_{\rm{BH}}-\sigma$ relation (see Fig. 1 of Shields et al. 2006). If we assume the 
smaller inclinations (10$^o$-15$^o$) for these narrow
CO quasars, by deriving the inclination-corrected  $\sigma$ value, we can see that they
can also fit the $M_{\rm{BH}}-\sigma$ relation. Shields et al. (2006) suspect that these
narrow CO quasars are starbursts confined to a small part of their host galaxies.
However, the possibility of smaller inclinations for these quasars can not be excluded.
 
In the above investigations we adopted the black hole masses for quasars given by 
Shields et al. (2006), who assumed random orbits of the broad-line region (BLR) clouds.
In fact, if the BLR is also in a disk like configuration, the BLR inclinations can also
affect the black hole mass estimations (McLure \& Dunlop 2001; Wu \& Han 2001; Collin 
et al. 2006). If we assume  a BLR  inclination angle {\it i} and a ratio {\it A} of the
random velocity component to the planar velocity component, the black hole mass 
increases by a factor of $1/3(sin^2i+A^2)$ (see Eq. (4) in Wu \& Han 2001), 
compared with that 
obtained by assuming a random orbit of BLR as done in Shields et al. (2006). If we assume that
the BLR inclinations of high-redshift quasars are around 15$^o$ and {\it A} is in the range
 0 - 0.3 (Collin et al. 2006), their black hole masses are larger by a factor of 
2 - 5 than the masses given by Shields et al. (2006). This means that the 
$M_{BH}-\sigma$ relation for the high-redshift quasars would be more offset from the 
local one if the CO line width was not corrected for inclination. However, if we assume that
the BLR inclination is about the same as the galactic inclination for high-redshift 
quasars and consider 
the inclination effects on both the black hole mass and bulge velocity dispersion 
estimations, we find that the $M_{BH}-\sigma$ relation for high-redshift quasars can still be
consistent with the local one if the inclination angles are around 10$^o$. Our conclusion 
is almost unchanged, even if the inclination effect on the
black hole mass is considered.  because  $M_{BH} \propto \sigma^4$, which indicates
that the inclination 
dependence of $\sigma^4$ ($\propto 1/sin^4i$) is much stronger than
that of $M_{BH}$ ($\propto 1/sin^2i$). 
For the low-redshift PG quasars, if their 
inclinations are around 30$^o$ - 45$^o$, the inclination will have less effect on their
$M_{BH}-\sigma$ relation, because the inclination corrections to both $M_{BH}$ and 
$\sigma$ are much smaller. Among 11 PG quasars with stellar velocity dispersions recently measured
by CO absorption features (Dasyra et al. 2006), seven  have detected CO 
emissions (Evans et al. 2001; Scoville et al. 2003; see also Shields et al. 2006). Using
the  $\sigma$ and FWHM  of CO emission-line data for these seven quasars, from Eq. (3) we 
can roughly estimate their inclination values to be in the range 20$^o$ - 45$^o$, 
consistent with 
that we assumed for PG quasars. However, we must note that the BLR geometry and dynamics
may not be as simple as we assumed. More accurate determinations of black hole masses, 
especially for high-redshift quasars, are still needed.  

\section {Summary and discussions}

We have presented an investigation of the 
correlation
between the CO line width and the bulge velocity dispersion using a sample of
33 nearby Seyfert galaxies. We found that the formula adopted in some previous 
studies, $\sigma=\rm{FWHM(CO)}/2.35$, is generally not a good approximation. 
Using it, one may significantly
underestimate the value of velocity dispersion  when the CO line 
is narrower
than 400 $km~s^{-1}$. By involving the galactic inclination angle  as an
additional parameter, we obtained a tight relation between the 
inclination-corrected CO line width 
FWHM(CO)/$\sin i$ and $\sigma$ (see Eq. (3)). Using this new relation, we can better
estimate the bulge velocity dispersion from the CO line width. We applied this new 
relation to nine high-redshift quasars 
with CO line detections and found that
they are consistent with the local $M_{BH}-\sigma$ relation if their 
inclination angles are around $15^o$. Therefore, we think that it is premature
to conclude that the high-redshift quasars do not follow the local 
$M_{BH}-\sigma$ relation based on a simple $\sigma$--FWHM(CO) relation without 
considering the effects of galactic inclinations. 

There are  three important assumptions in our study. One is that we assume that the CO emissions
are not uniform and are mainly concentrated in a rotating galactic disk. This is probably
true because the molecular gas has been seen in such a configuration at subkiloparsec
scales in most 
low-redshift ultraluminous infrared galaxies (Downes \& Soloman 1998). In addition,
the double-peaked CO line profiles have been frequently seen in Seyfert galaxies,
low-redshift quasars and SMGs (Maiolino et al. 1997; Evans et al. 2001; Greve et al. 
2005). These profiles are likely to be produced if the  molecular gas is in a
rotating disk, although such an explanation is not unique (Carilli \& Wang 2006). 
The second assumption we have made is the smaller inclinations of the high-redshift 
quasars. The smaller inclinations (that is, a face-on geometry) of these
high-redshift quasars  are indeed preferred because we  would be unable to
detect them if they were viewed edge-on. Moreover, the smaller inclinations of these
high-redshift quasars are
also consistent with their relatively smaller CO line
widths in comparison with other SMGs at high redshift
having similar total amount of molecular gas (Greve et al. 2005; Carilli \& Wang 2006),
although we can not exclude other possibilities for the narrowness of CO lines of
these quasars, such as the different stages of the merge sequence or galaxy mass or size
(see discussions in Carilli \& Wang 2006).  The third assumption is that
the CO dynamics for the local Seyfert galaxies and high-redshift quasars are not 
dramatically different. This obviously needs
to be confirmed by future studies. The future Atacama Large Millimeter Array (ALMA)
telescope will have both high sensitivity and resolution, and will be able to
easily reveal the CO dynamics of high-redshift quasars.  

In order to get a better understanding of the $M_{BH}-\sigma$ relation at high redshift,
we have to rely on more accurate measurements of both $M_{BH}$ and $\sigma$ for high-redshift quasars and galaxies in the future. The CO imaging and spectroscopic studies on
these distant objects with more advanced future instruments will tell us more about the
geometry and dynamics of the molecular gas, as well as their relation to the galactic 
bulge. These studies are absolutely needed to check the relationship between the
 CO line width and the bulge velocity dispersion we derived here. In addition, much work is needed to do to get better estimates of the black hole masses for
distant quasars using various more advanced techniques. These efforts will undoubtedly greatly
improve our knowledge about the black hole -- bulge relation in the early universe.

\acknowledgments
 I thank Chris Carilli, Yu Gao,  Fukun Liu, Ran Wang and Bingxiao Xu for stimulating
 discussions  and the anonymous referee for helpful suggestions. This work
is supported by NSFC grants (10473001 and 10525313), an RFDP
grant (20050001026), and the Key Grant Project of the Chinese Ministry
of Education (305001). I acknowledge use of the HyperLeda database 
(http://leda.univ-lyon1.fr).

\clearpage
\begin{figure}
\centering
\plotone{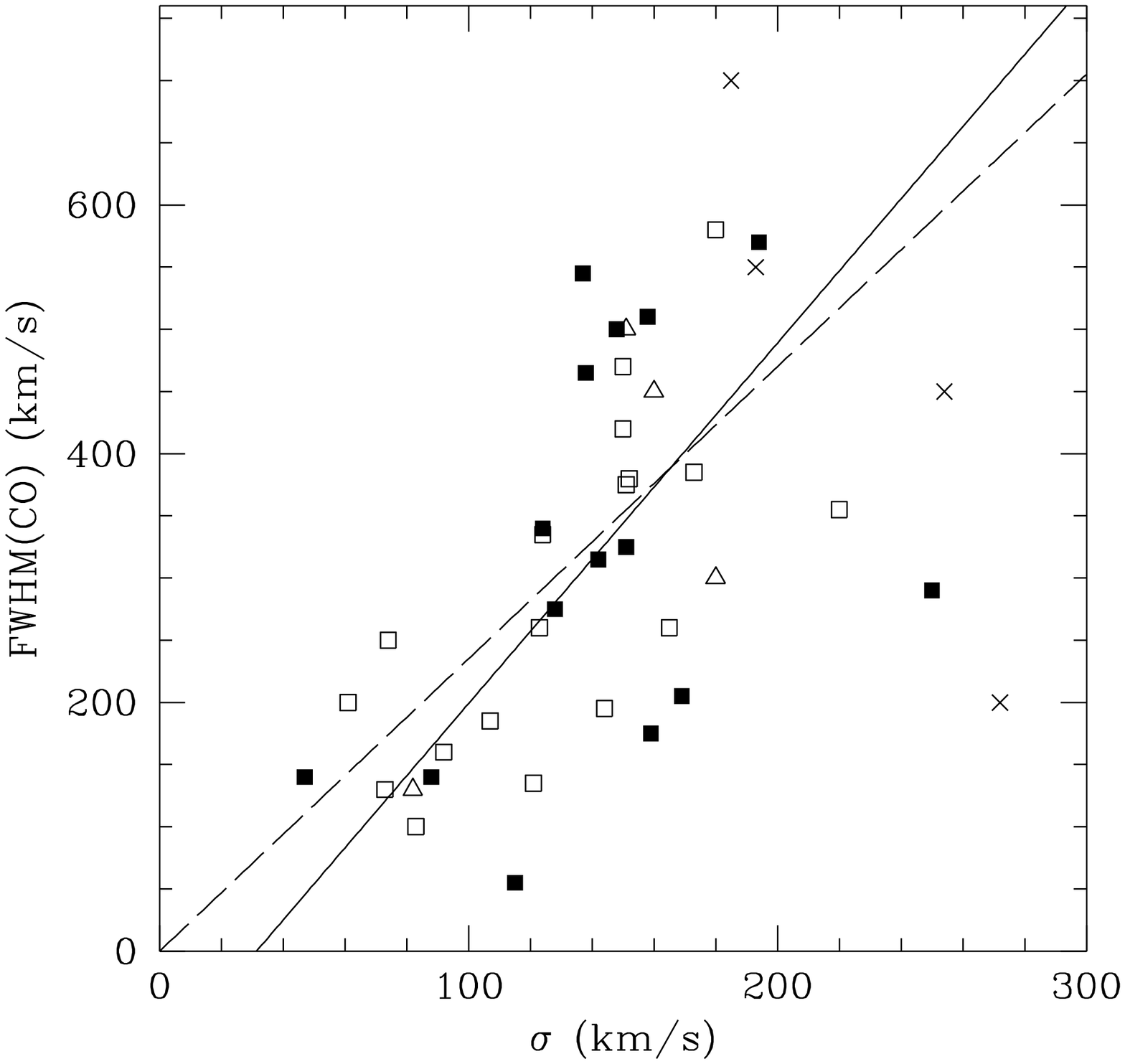} \caption{The relation between the CO line width and the 
bulge stellar velocity
dispersion.  The filled and open squares represent
Seyfert 1s and Seyfert 2s in our main sample. The triangles represent 4 
additional Seyferts in 
Heckman et al. (1989) and the crosses represent 4 radio galaxies in 
Evans et al. (2005).
The solid line represents the OLS besector fit. The dashed line shows 
$\sigma$=FWHM(CO)/2.35.
\label{f1}}
\end{figure}

\clearpage
\begin{figure}
\centering
\plotone{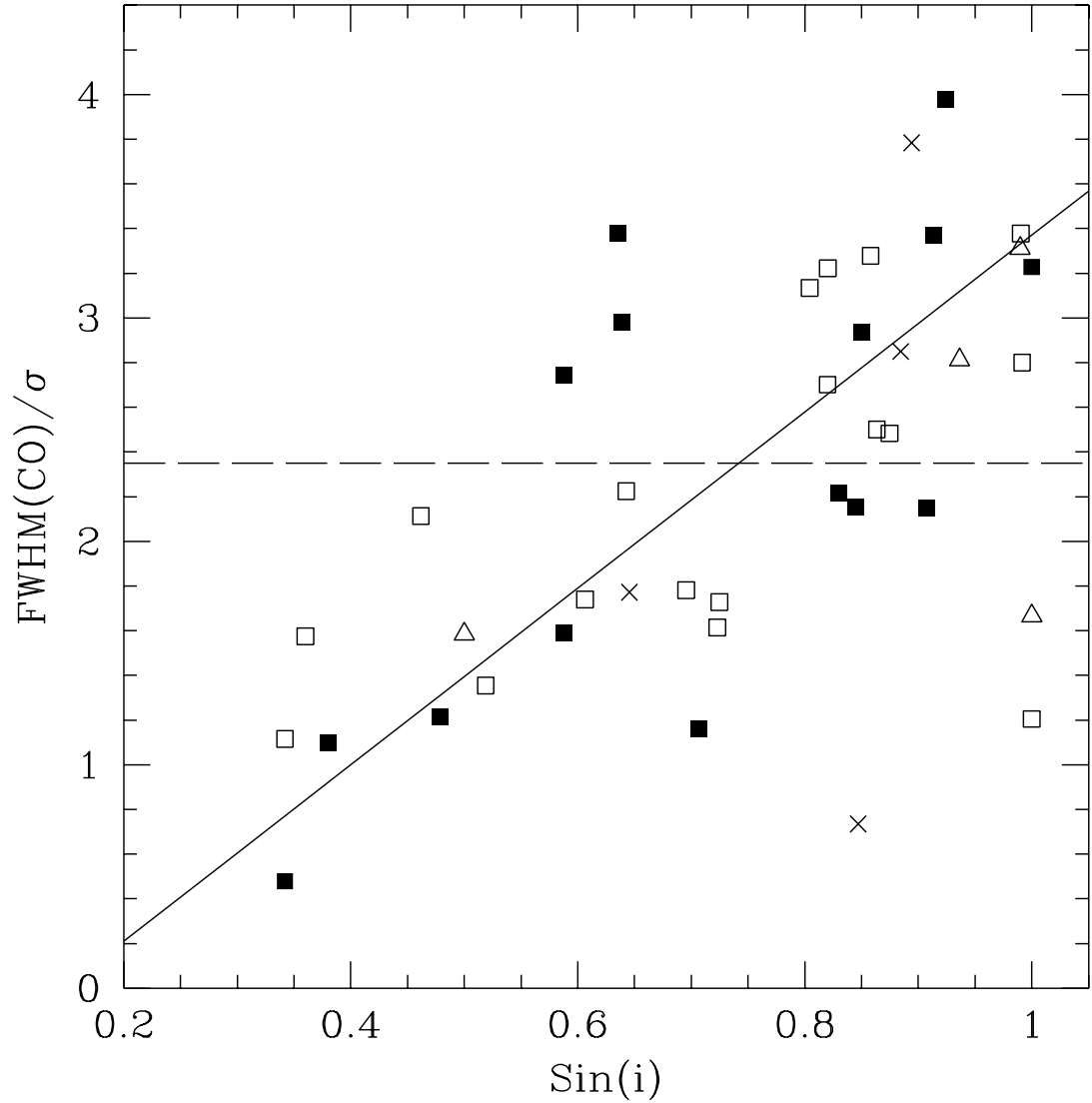} \caption{The relation of the ratio of CO line width and 
the bulge velocity
dispersion with the galactic inclination. The solid line shows 
the OLS bisector fit. The dashed line shows 
$\sigma$=FWHM(CO)/2.35. The symbols 
have the same meanings as in Fig. 1.
\label{f2}}
\end{figure}

\clearpage
\begin{figure}
\centering
\plotone{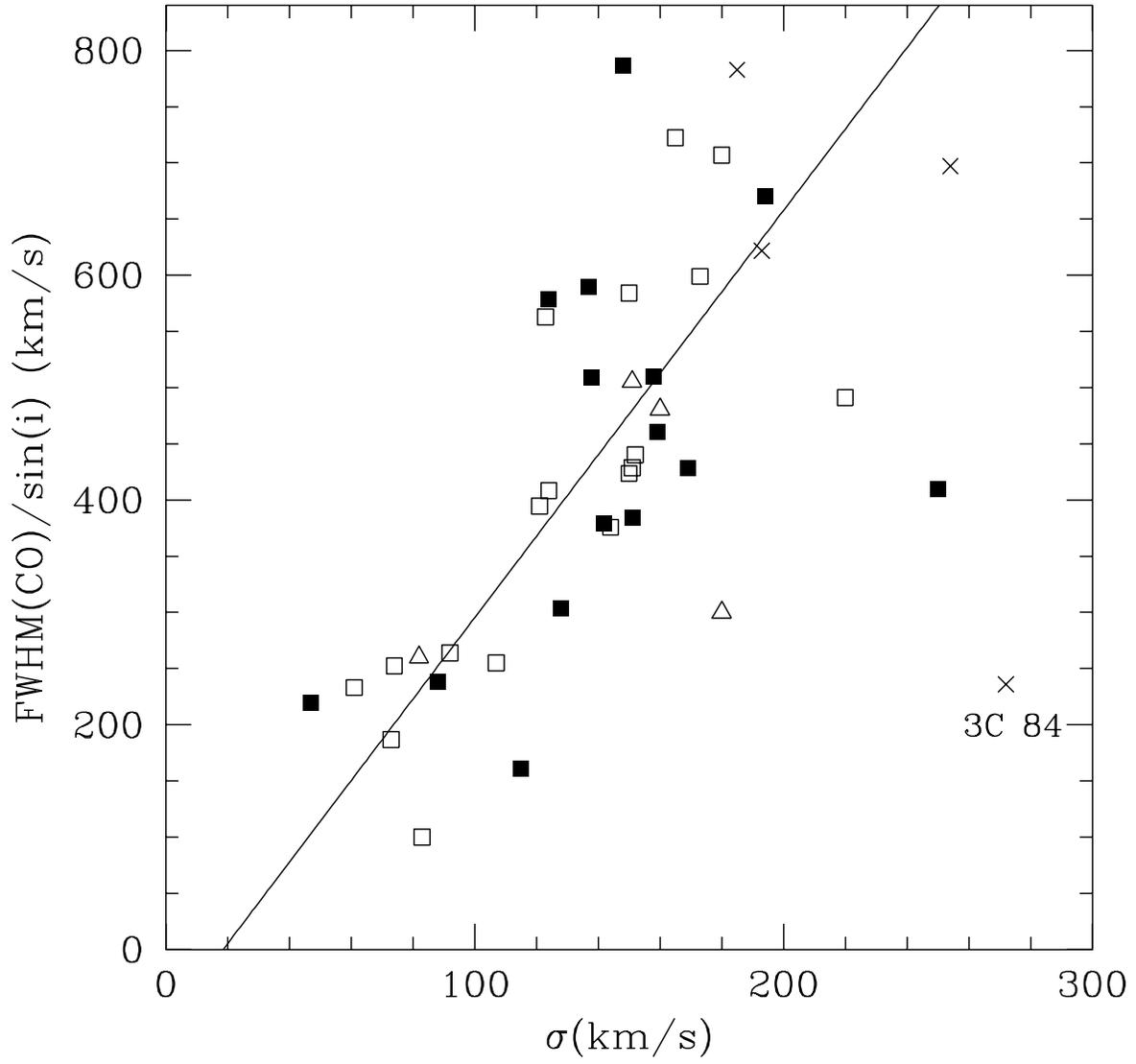} \caption{The relation between the inclination-corrected
 CO line width and 
the bulge velocity
dispersion.  The solid line shows 
the OLS bisector fit. The symbols 
have the same meanings as in Fig. 1. The extreme outlier object 3C 84 is
indicated.
\label{f3}}
\end{figure}

 \clearpage
\begin{figure}
\centering
\plotone{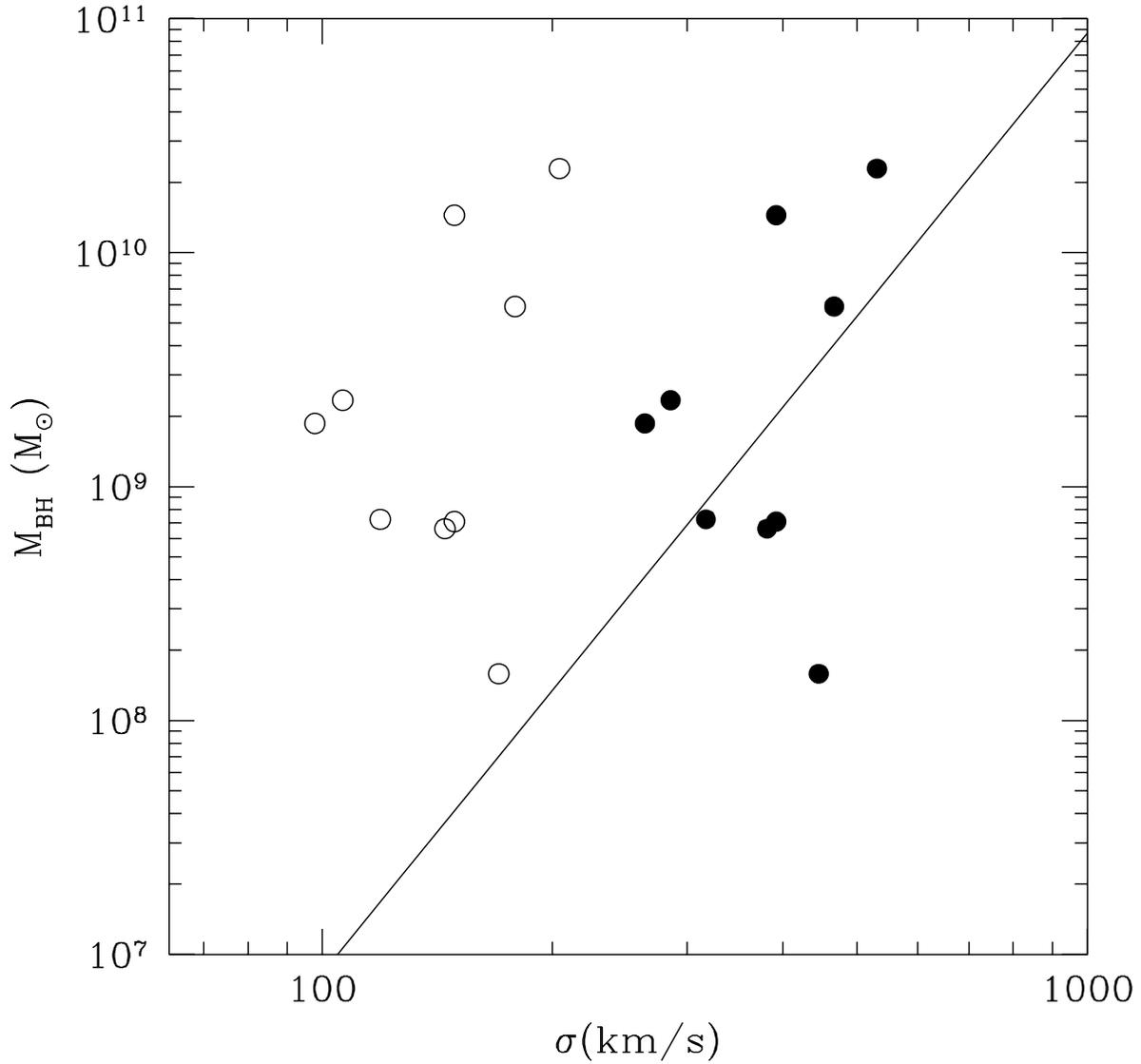} \caption{The $M_{\rm{BH}}-\sigma$ relation for high 
redshift quasars. The open circles correspond to  $\sigma$ values derived
from $\sigma$=FWHM(CO)/2.35, while the filled circles correspond to  
$\sigma$ values derived from the inclination-corrected CO line width by
assuming the inclination of 15$^o$. The solid line shows the local
 $M_{\rm{BH}}-\sigma$ relation given by Tramaine et al. (2002).
\label{f4}}
\end{figure}

\clearpage

\begin{deluxetable}{lccccccc}
\tabletypesize{\scriptsize} \tablewidth{0cm} \tablecaption{The data of 33 
Seyferts in our main sample and 8 additional AGNs\label{tbl-1}}
\tablehead{\colhead{Name} & \colhead{Type$^a$} & \colhead{z} & \colhead{FWHM(CO)}
 & \colhead{Ref$^b$} & \colhead{$\sigma$} & \colhead{Ref$^c$} & \colhead{i$^d$}\\
& & & ($km~s^{-1}$) & &($km~s^{-1}$)& &(degree)}\startdata
NGC 513 & S2 & 0.0195 & 380 & M97 & 152$\pm$10 & NW95 & 59.7\\ 
Mrk 573 & S2 & 0.0169 & 260 & M97 & 123$\pm$16 & NW95 & 27.5\\
NGC 863 & S1 & 0.0263 & 205 & M97 & 169$\pm$28 & NW95 & 28.6\\
NGC 1068 & S2 & 0.0038 & 260 & M97 & 165$\pm$12 & NW95 & 21.1\\
NGC 1365 & S1 & 0.0054 & 325 & M97 & 151$\pm$20 & OOKM95 & 57.7\\
NGC 1667 & S2 & 0.0151 & 385 & M97 & 173$\pm$17 & TDT & 40.0\\
NGC 2110 & S2 & 0.0076 & 355 & M97 & 220$\pm$25 & NW95 & 46.3\\
NGC 2273 & S2 & 0.0062 & 335 & M97 & 124$\pm$10 & NW95 & 55.1\\
 Mrk  10 & S1 & 0.0292 & 545 & M97 & 137$\pm$38 & NW95 &67.5\\
 NGC 2992 & S1& 0.0077 & 510 & M97 & 158$\pm$13 & NW95 &90.0\\
 Mrk 1239 & S1& 0.0196 & 290 & M97 & 250$\pm$20 &OOMM99 & 45.0\\
 NGC 3079 & S2& 0.0037 & 420 & M97 & 150$\pm$10& SWC93 & 82.5\\
 NGC 3185 & S2& 0.0041 & 200 & M97 & 61$\pm$20 &NW95 &  59.1\\
 NGC 3227 & S1& 0.0038 & 275 & M97 & 128$\pm$13 &NW95 & 65.1\\
 NGC 3362 & S2& 0.0277 & 160 & M97 & 92$\pm$28 &NW95 &  37.3\\
 NGC 3786 & S1& 0.0090 & 315 &M97 & 142$\pm$13&  NW95 &56.1\\
 NGC 4051 & S1& 0.0023 & 140 &M97 & 88$\pm$13 &NW95& 36.0\\
 NGC 4388 & S2& 0.0084 & 250 &M97 &74$\pm$34 &NW95&82\\
 NGC 4501 & S2& 0.0076 & 375 &M97 &151$\pm$17& HS98&61\\
 NGC 4593 & S1& 0.0083 & 340 & M97 & 124$\pm$29 & NW95&36.0\\
 NGC 5033 & S1& 0.0029 & 465 & M97 & 138$\pm$10 &HS98 &  66.0\\
 NGC 5347 & S2& 0.0079 & 130 & M97 & 73$\pm$14 & NW95&  44.1\\
 NGC 5548 & S1& 0.0171 & 140 & M97 & 47$\pm$67 & NW95& 39.7\\
 NGC 5929 & S2& 0.0085 & 135 & M97 & 121$\pm$13&  NW95&  20.0\\
 NGC 5940 & S1& 0.0340 & 175 & M97 & 159$\pm$49& NW95& 22.3\\
 NGC 5953 & S2& 0.0065 & 185 & M97 & 107$\pm$8&  NW95& 46.5\\
 Arp 220 &  S2& 0.0181 & 470 & M97 & 150$\pm$4 & SF-98 & 53.6\\
 NGC 6104& S1&  0.0279 & 500 & M97 & 148$\pm$35& NW95&  39.5\\
 NGC 6814& S1&  0.0052 &  55 & M97 & 115$\pm$18& NW95 & 20.0\\
 NGC 7172& S2&  0.0086 & 580 & M97 & 180$\pm$40&  Lon+94b &55.1\\
 Mrk 530 & S1&  0.0293 & 570 & M97 & 194$\pm$31 & NW95 & 58.2\\
 NGC 7674& S2&  0.0295 & 195 & M97 & 144$\pm$32 &   NW95& 31.3\\
 NGC 7743& S2&  0.0057 & 100 & M97 & 83$\pm$20&  NW95&  90.0\\
\hline
NGC 931&  S1& 0.0164 & 500 & H89 & 151$\pm$32 & NW95 & 81.8\\
M51    &  S2& 0.0016 & 130 & H89 & 82$\pm$11 & NW95 & 30.0\\
NGC 5506& S2& 0.0058 & 300 & H89 & 180$\pm$20 & OOMM99 & 90.0\\
Mrk 273 & S2& 0.0373 & 450 & H89 & 160$\pm$60 & Jam+99 & 69.5\\
\hline
3C 31 & RG&0.0169 & 450 & E05 & 254$\pm$22 & SHI90 & 40.2\\
3C 84 & RG&0.0176 & 200 & E05 & 272$\pm$61 & NW95 & 57.9\\
3C 120 &RG&0.0331 & 550 & E05 & 193$\pm$40 &  SHI90 &62.2\\
3C 293 &RG&0.0448 & 700 & E05 & 185$\pm$20 & H+85 &63.4\\
\hline 
\enddata
\tablecomments{a: Types of AGN. S1, S2 and RG represent Seyfert 1, 
Seyfert 2 and radio galaxy, respectively.
b: Refereces for the CO line width data. M97: Maiolino et al. (1997);
H89: Heckman et al. (1989); E05: Evans et al. (2005).
c: References for the bulge velocity dispersion data. NW 95: Nelson \&
Whittle (1995); OOKM95: Oliva et al. (1995); OOMM99: Oliva et al. (1999); 
TDT: Terlevich, Diaz \& Terlevich (1990); 
SWC93: Shaw, Wilkinson \& Carter (1993); HS98: Heraudean \& Simien (1998); 
SF-98: Shier \& Fisher (1998);  Lon+94b: Longo et al. (1994); Jam+99: James et al. (1999);
SHI90: Smith, Heckman \& Illingworth (1990); H+85: Heckmann et al. (1985).
d: inclination data are from the Hyperleda database except for NGC 6814, whose
inclination is taken from Whittle (1992).}
\end{deluxetable}

\end{document}